**Magnetism of fine particles of Kondo lattices, obtained by high-energy ball-milling**

E.V. Sampathkumaran, K. Mukherjee, Kartik K Iyer, Niharika Mohapatra and Sitikantha D. Das
Tata Institute of Fundamental Research, Homi Bhabha Road, Colaba, Mumbai-400005, India.

E-mail: sampath@mailhost.tifr.res.in

**Abstract:** Despite intense research in the field of strongly correlated electron behavior for the past few decades, there has been very little effort to understand this phenomenon in nano particles of the Kondo lattices. In this article, we review the results of our investigation on the fine particles ($<< 1\ \mu m$) of some of the alloys obtained by high-energy ball-milling, to bring out that this synthetic method paves a way to study strong electron correlations in nanocrystals of such alloys. We primarily focus on the alloys of the series, $CeRu_{2-x}Rh_xSi_2$, lying at different positions in Doniach's magnetic phase diagram. While $CeRu_2Si_2$, a bulk paramagnet, appears to become magnetic (of a glassy type) below about 8 K in fine particle form, in $CeRh_2Si_2$, an antiferromagnet ($T_N$= 36 K) in bulk form, magnetism is destroyed (at least down to 0.5 K) in fine particles. In the alloy, $CeRu_{0.8}Rh_{1.2}Si_2$, at the quantum critical point known to exhibit non-Fermi liquid behavior in the bulk form, no long range magnetic ordering is found.

**1. Introduction**

Though there has been considerable emphasis to understand the physics and chemistry of solids in nano dimensions, very little efforts have been made to probe strongly correlated electron systems of rare-earth ($R$) alloys in such small particles. During last ten years or so, there were a few attempts only [1] in this direction, e.g., on a few binary Ce alloys, $CeAl_2$ and $CePt_2$, prepared by a flash-evaporation method in very small quantities. It was reported that the Kondo effect is altered at the expense of magnetic ordering in these alloys. Clearly, the progress in the direction of 'strong electron correlation in nano specimens' of Kondo lattices is rather slow. This is presumably due to the difficulties in the synthesis of large quantities of nano particles of such rare-earth alloys in a stable form with an ease to handle in air to facilitate various measurements. The purpose of this article is to bring out that the high-energy ball milling in a medium of toluene offers a route to synthesize stable forms of such fine particles in a large quantity.

High-energy ball-milling is a well-established mechanical method employed by materials engineers to synthesize nano particles [2]. In fact, nano magnetism of binary Gd and Tb alloys in reduced dimensions was studied [3] on the specimens prepared by this method. During the course of our investigations, similar studies were reported on $YbAl_3$ [4]. In recent years, we have carried out substantial investigations on ball-milled specimens of rare-earth intermetallics, e.g., $RCo_2$ [5, 6], $RMn_2Ge_2$ [7], $RCr_2Si_2C$ [8]. One of the important outcomes was that the exchange-enhanced Pauli paramagnets (in bulk form), like $YCo_2$, $LuCo_2$, and $ZrCo_2$, which have been known to undergo itinerant electron metamagnetism at very high fields (>500 kOe), are found to become ferromagnetic in nano form at room temperature [5, 6]. With this clue, we wanted to probe how a heavy-fermion with a non-magnetic ground state (in bulk) behaves in the nano form far away from QCP as well as at QCP. We therefore investigated the alloys, $CeRu_{2-x}Rh_xSi_2$, crystallizing in $ThCr_2Si_2$-type tetragonal structure. The compound, $CeRu_2Si_2$, is a well-known non-magnetic Kondo lattice far away from QCP [9, 10], while $x=1.2$ alloy is at QCP [11]. Earlier [6], on the basis of initial magnetization studies, we reported that, $CeRu_2Si_2$ undergoes some kind of magnetic ordering in the fine particles of this compound. However, in the latter (reported here), there is no onset of magnetic ordering. As a natural extension of this work, we further studied $CeRh_2Si_2$ ($T_N$= 36 K in the bulk form, Ref. 12) as well and we find that magnetic ordering is suppressed (down to 0.6 K) in the fine particles of this compound. All the results are summarized

## 2. Experimental details

The polycrystalline ingots of the compositions, $x=$ 0, 1.2, and 2.0, of the series $CeRu_{2-x}Rh_xSi_2$, were prepared by arc melting stoichiometric amounts of high-purity (>99.9%) constituent elements in an arc furnace in an atmosphere of argon. The bulk samples thus obtained were characterized by x-ray diffraction (XRD) (Cu $K\alpha$); magnetic measurements were further carried out and the data were compared with those known in the literature to ascertain the quality of the bulk specimens. The materials were then milled in a planetary ball-mill (Fritsch pulverisette-7 premium line) operating at a speed of 500 rpm in a medium of toluene for different intervals of time. Zirconia vials and balls of 5mm diameter were used and the ball-to-material ratio was kept at 5:1. These vials and balls contain 96.4% of $ZrO_2$, and 3.2% of MgO and thus possible contamination from magnetic impurities due to these milling parts can be ignored. [We have prepared specimens with tungsten carbide vials and balls as well. Though major results were not influenced, traces of Co impurity from these vials were found to distort absolute values of magnetization (*M*) marginally]. XRD patterns confirm single-phase nature of all specimens, the scanning electron microscopic (SEM, Nova NanoSEM600, FEI Company) pictures established homogeneity, and the energy dispersive x-ray analysis (EDXA) confirmed that the atomic ratios of the constituent elements are the same as those of ingots. The same SEM was employed to confirm the reduction of particles to nanometer size. In some cases, a transmission electron microscope (TEM, Tecnai 200 kV) was also employed to look at particle sizes and selected area diffraction pattern was also obtained to confirm the structure (see figure 2). The XRD patterns (along with the average particle sizes determined from Debye-Scherrer formula) and SEM pictures are shown in figure 1 and 2 respectively. While the widths of x-ray diffraction lines reveal that the particle sizes fall well below 50 nm, SEM pictures show significant agglomeration of these particles. At this juncture, we would like to mention that we attempted to anneal the milled specimens at about 100 C (above the boiling point of toluene) for 5 minutes in evacuated sealed quartz tubes to remove strains, but the samples tend to deteriorate on exposure to air (as evidenced by XRD patterns). This implies that toluene coating exists and it plays a role on the stability of the milled (and dried) specimens. The *dc* magnetization (in the range 1.8-300 K), *ac* susceptibility *($\chi$)* and *M(H)* (up to 50 kOe) measurements at selected temperatures for all specimens were carried out with the help of a commercial superconducting quantum interference device (Quantum Design). In addition, in some cases, wherever there is a need to extend *M(H)* to higher fields, we have employed a commercial vibrating sample magnetometer (Oxford Instruments). We have also performed heat capacity (*C*) measurements on the fine particles employing a physical property measurements system. For this purpose, we first measured *C(T)* of a known quantity (a few mg) of dried GE varnish and the values of *C* were found to be negligible. Therefore, we mixed a known mass of the fine particles in GE varnish, dried and used this for measuring heat-capacity. We find that this method of specimen-preparation offers a reliable way to measure heat-capacity of nano specimens.

## 3. Results

**CeRu$_2$Si$_2$:**

For this sample, we did not see any significant change in the diffraction-line intensities as the milling-time is increased from two and half hours to 6 hours (not shown here). It appears that milling well beyond 30 minutes does not significantly change the particle size in this compound. Thus, it is inferred that one could obtain fine-particles within few hours of milling. We report the data for 2.5 h milled sample.

In our previous publication [6], we already discussed in detail the magnetization behavior of the specimens obtained by milling using tungsten carbide vials and balls. Since the features observed for the present freshly prepared specimens are the same as in that publication, we present here essential features based on fresh measurements. As inferred from figure 3***a***, there is a sudden enhancement of $\chi$ value (measured in 5 kOe) around 8 K for the milled specimens; we reported that the $\chi(T)$ curves obtained for the zero-field-cooled (*ZFC,* from 35 K) and field-cooled (FC) conditions of the specimens tend to bifurcate at this temperature [6]. There are corresponding anomalies in the *ac* $\chi(T)$ data as well ($H_{ac}$= 1 Oe) for both real and imaginary parts (not shown here, see Ref. 6). Though these results establish the existence of a magnetic transition near 8 K, we could not confirm from these data whether the ordering is of a ferromagnetic type or a spin-glass type. We have also measured isothermal *M* behavior at 1.8 K up to 120 kOe to throw more light on magnetism (see figure 3***b***). *M(H)* behavior of the bulk form is in conformity with that known for polycrystals in the literature [Besnus et al, Ref. 9] establishing spin reorientation around 70 kOe. Notably, for the milled-specimen, the nature of *M(H)* curve is completely modified in the sense that the spin-reorientation feature disappears. The fact that the *M(H)* curve is (weakly) hysteretic implies a ferromagnetic component in the fine-particles. However, there is no evidence for saturation of *M* till 120 kOe. The *C(T)* curve (figure 1***c***) reveals a prominent broad shoulder below 10 K indicative of a magnetic transition (but with a well-defined magnetic structure), which was absent in bulk form. We would like to reiterate that the value of the magnetic-moment obtained at high fields, say at 120 kOe (about 0.5 $\mu_B$/Ce) is significant enough to conclude that the observed features and inferences can not be attributable to surface or any impurity [13]. Finally, the plots of inverse $\chi$ versus *T* for the bulk and the milled specimens overlap obeying Curie-Weiss law above ~125 K as shown in figure 3***d***. The value of the effective moment (~2.4 $\mu_B$/Ce) obtained from the Curie-Weiss region turns out to be very close to that expected for trivalent Ce ion. These findings rule out deterioration (e.g., oxidation) of the samples due to milling (though a modification of the curve following milling does not necessarily imply instability of the compound). The paramagnetic Curietemperature ($\theta_p$) is about 38 K as in our bulk sample. In short, in this non-magnetic Kondo lattice, ball-milled nano specimens appear to exhibit magnetic ordering of a complex type at low temperatures.

**$CeRu_{0.8}Rh_{1.2}Si_2$:**

It is now established [11] in the literature that a partial replacement of Ru by Rh drives Ce towards magnetic ordering in the bulk form, with $x= 1.2$ alloy lying at the quantum critical point, exhibiting non-Fermi liquid behavior (NFL) in its properties (for example, logarithmic divergence of *C/T* at low temperatures). In view of the observation of magnetic ordering in the nano form of $CeRu_2Si_2$, it is important to see whether the same is observed at QCP. In fact, during the course of this investigation, Kim et al [14] reported magnetic properties of this alloy with varying particle size and they found Griffiths-like behavior only without any evidence for any well-defined magnetic transition temperature. In order to enable us a comparison with the report of Kim et al [14], we present our findings on two specimens, one after 2 hour milling and another after 5 hour milling. As stated earlier, the average particle size determined from XRD pattern (say, from the line-width of the most intense line) is found to fall in the range of 20 to 30 nm with a marginal decrease beyond 2 h milling. In figure 4, we show *M(T)* as a function of *T* (above 2 K) in different ways. There is no feature attributable to the onset of magnetic ordering in the entire temperature range of investigation (see figure 4**a**, the curves below 40 K) for both the specimens. This was confirmed by ac χ data as well, in which real part was found to exhibit a monotonic increase with decreasing temperature (not shown). DC χ exhibits Curie-Weiss behavior at high temperatures (>200 K) with an effective moment very close to that expected for trivalent Ce ion even for the milled specimens, with a gradual increase of $\theta_p$ as though the high-temperature Kondo-temperature increases as the particle size is reduced; this trend of $\theta_p$ appears to be reversed at low temperature (see figure 4b). At this juncture, we would like to remark that Kim et al [14] reported a dramatic enhancement of dc χ with decreasing particle size, but we do not find this behavior above 10 K in our studies. This might mean that either this finding of Kim et al [14] is not intrinsic or χ exhibits a peak as a function of particle size with a gradual fall below a critical particle size. If the latter explanation is valid, this is an interesting behavior. However, below 10 K, our results are in agreement with that of Kim et al [14] (see figure 4*a*). Now, turning to isothermal magnetization behavior measured at 1.8 K (see figure 4*c*), *M* varies essentially linearly with *H* for the bulk (below 50 kOe), whereas in the milled specimens, it varies nearly as $H^{\sim 0.65}$, with the exponent very close to that reported by Kim et al [14] who attributed it to Griffiths phase behavior. Finally, we show the heat-capacity data above 0.6 K for the milled specimens in figure 4*d* and we do not find the broad shoulder around 8 K noticed for $CeRu_2Si_2$ in this alloy, but C monotonically decreases with *T*. A plot *C/T* vs. T (Fig. 4*d*, inset) reveals a logarithmic component at low temperatures with the absolute values increasing with milling time (in other words, with decreasing particle size), similar to the finding of Kim et al [14], though we could not compare the trend below 0.6 K. In short, the present study establishes that this alloy at QCP does not become magnetically ordered with a reduction in size to nanometric dimensions.

**CeRh$_2$Si$_2$:**

We now focus on the comparison of the magnetic behavior of this compound - which is at the peak [15] of Doniach's magnetic phase diagram – in the bulk and in fine particles. We have carried out detailed magnetic studies for the specimen milled for 5 hours only. In figure 5*a* and 5*b*, we show *C* and *M* data as a function of temperature. It is straightforward from this figure that the feature at about 36 K due to magnetic ordering of Ce sublattice is completely destroyed in both the plots and there is no evidence for magnetic ordering (down to 0.5 K, extended in the case of heat capacity). The plots of *C/T* versus $T^2$ (Fig. 5*c*) reveal that the linear coefficient of heat capacity from the region 5-15 K is about 100 mJ/molK$^2$ for the fine particles and it increases to about 200 mJ/mol K$^2$ at 2 K. These indicate that this compound in the nano form becomes a heavy fermion with a loss of magnetic ordering from Ce. We however note an interesting finding in the plot of inverse χ versus *T* (Fig. 5d). This plot is linear above 100 K in both bulk and nano form, but the effective moment obtained from the linear region is marginally higher (~ 3μ$_B$) for the fine particles than that expected for trivalent Ce ion unlike in bulk compound. The value of θ$_p$ also gets enhanced from about -60 K to -320 K. It is not clear whether Rh acquires a magnetic moment in the milled specimen or whether there is a formation of another magnetic impurity in traces undetectable by X-ray or SEM. It may be remarked that we see a sudden change in slope around 60 K (as marked by asterisk in figure 5*d*) for the fine particles, and it is not clear whether this arises from itinerant magnetic ordering from Rh sublattice. Future experiments may focus on this issue. With respect to isothermal *M* behavior (Fig. 5*e*), the plot is almost linear till 120 kOe for the bulk specimen even at 1.8 K. For the milled specimen, the values of the magnetic moment are enhanced at high fields with respect to that for the bulk; while the plots remain linear till 120 kOe say, at 20 K (not shown), a curvature develops as the temperature is lowered as if there is a tendency for saturation. It is possible that this curvature arises from traces of uncompensated Ce leading to magnetism. A careful look at the plot of *C/T* versus *T* at low temperatures reveals that there are some weak structures around 1 K (see inset of figure 5*c*). Nevertheless, the results reveal that vast amount of Ce is driven non-magnetic for this compound in the fine particles.

**4. Discussion**

From the above results, it is clear that, in a Kondo lattice, CeRu$_2$Si$_2$, with a non-magnetic ground state (in bulk form) far above QCP in Doniach's magnetic phase diagram, magnetic ordering is induced in the Ce sublattice in fine particles. It may be recalled that the 4f-localization due to lattice expansion caused by a small replacement of Si by Ge or Ce by La in CeRu$_2$Si$_2$ results in a magnetic transition in the same temperature range. It thus appears that a reduction in particle size by milling essentially mimics the lattice expansion behavior in this structure, similar to that reported in the Mn-based series [7]. Extrapolating this "4f-localization" argument to CeRh$_2$Si$_2$, the bulk form of which is at the peak of Doniach's magnetic phase diagram [15], one would expect a suppression of magnetism for this compound in the nano form (due to weakening of 4f-conduction band coupling strength resulting from further 4f-localization) which is found to be the case experimentally. Similar significant weakening of 4f-conduction-band coupling could be responsible for the non-observation of bulk magnetic ordering for the *x* = 1.2 composition. We have extended [16] these studies to other magnetically ordering Kondo lattices in the same structure and we find a similar suppression of magnetism in $CePd_2Si_2$ and $CeAu_2Si_2$, consistent with the trends observed in the Ru(Rh)-based alloys. However, we do not find any change in the peak positions of lines in the XRD patterns with milling, implying that there is no change in the mean lattice constants. Therefore, lattice volume change is not the origin of observed features. This raises a question whether the electronic structure changes following reduced dimension and/or defects play a role in deciding the magnetic characteristics of milled specimens. At this juncture, it may be recalled that the role of defects on altering electronic density of states (and hence magnetism) was addressed in the case of binary Laves phase compounds [17] and therefore a role of such a factor in ball-milled specimens can not be ruled out.

## 5. Conclusion

We have synthesized the nano particles of some of the Ce-based alloys in a single series, crystallizing in $ThCr_2Si_2$-type tetragonal structure, by high-energy ball-milling and studied their magnetic behavior. The primary aim of this work is to bring out a method in which large quantities of fine particles of Kondo lattices can be synthesized in stable form to enable further investigations. We must however admit that, at present, this method does not allow us to synthesize fine particles of uniform size; also, the role of strain on the properties is not clear. Despite these deficiencies, we hope that the proposed route of synthesizing nano particles would serve as a starting point to synthesize nano particles of strongly correlated f-electron systems. Thus this work may serve as a triggering force to intensify research in this direction, as an attempt to reduce dimensions appear to change the properties under favorable circumstances whatever be its origin in detail.

The authors would like to acknowledge N.R. Selvi and G.U. Kulkarni, Jawaharlal Nehru Center for Advanced Scientific Research, Bangalore, India, for FESEM measurements and B. A. Chalke for TEM studies. We would like to thank Dr. J. Surya Narayana Jammalamadaka for his participation in experiments during initial phase of this work.

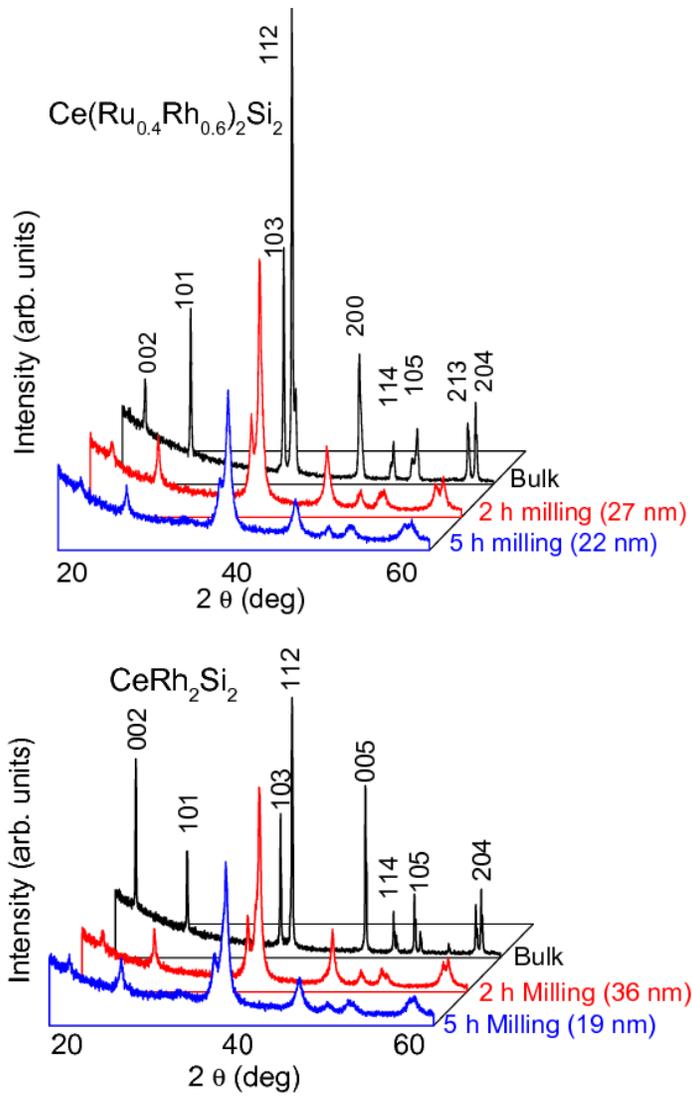

Figure 1:
X-ray diffractions of bulk and milled specimens of $CeRu_{0.8}Rh_{1.2}Si_2$ and $CeRh_2Si_2$. The particle sizes estimated by Debye-Scherrer formula from the half-width of the most intense line are indicated.

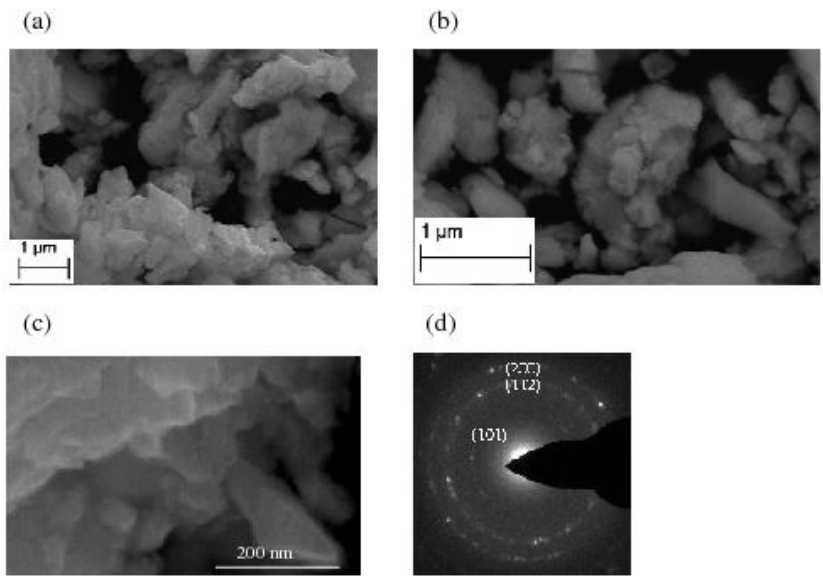

Figure 2:

Scanning electron microscopic pictures of the (final) milled alloys, (a) $CeRu_{0.8}Rh_{1.2}Si_2$. (b) $CeRh_2Si_2$, and (c) $CeRu_2Si_2$. Selected area diffraction pattern of the final specimen of $CeRu_2Si_2$ is also shown to establish proper phase formation of the nano specimen.

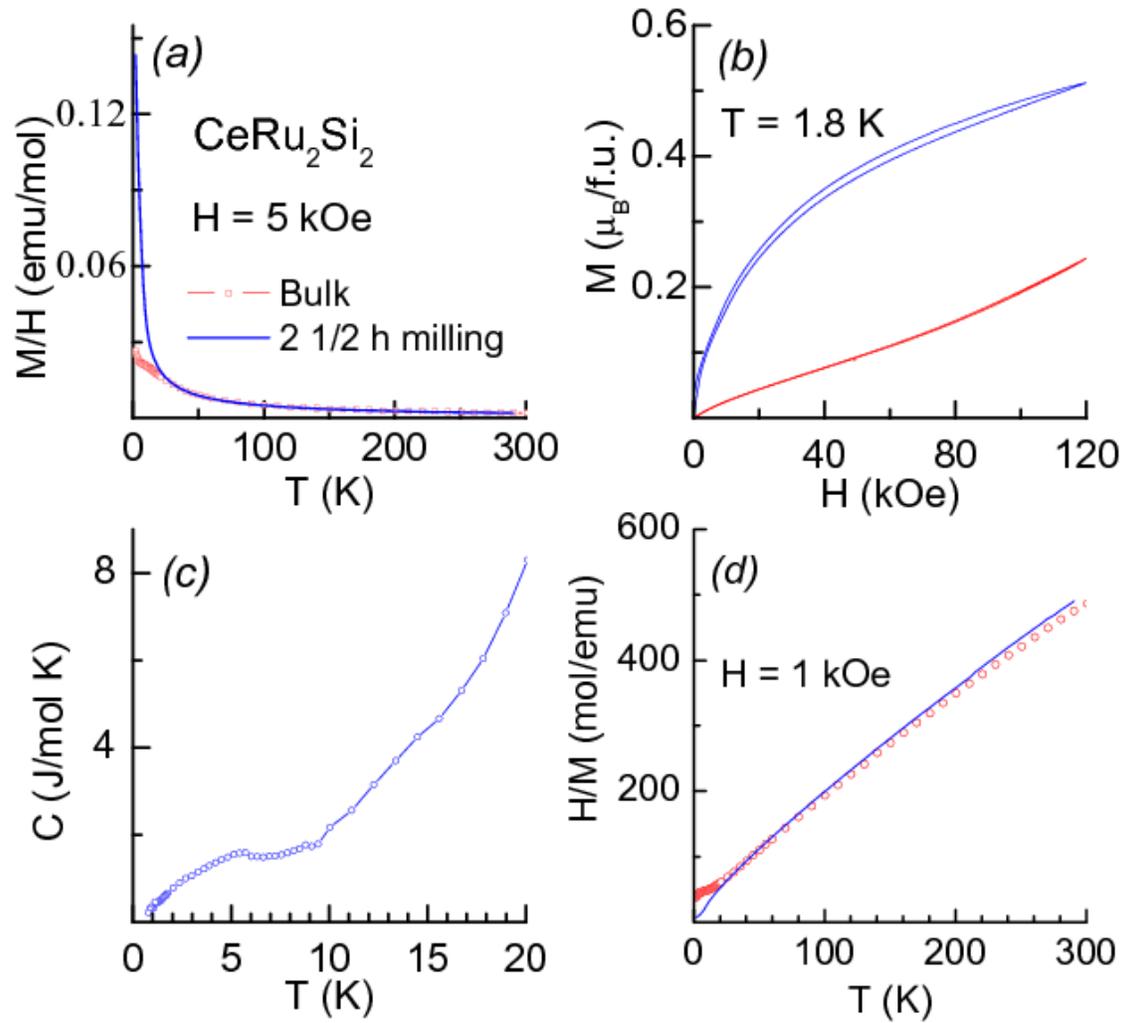

Figure 3:

(a) Magnetization divided by magnetic field as a function of temperature, *(b)* isothermal magnetization as a function of magnetic field at 1.8 K, *(c)* heat capacity as a function of temperature, and *(d)* inverse magnetic susceptibility as a function of temperature measured in a magnetic field of 1 kOe, for $CeRu_2Si_2$. The line through the data points serve as a guide to the eyes in *(c).*

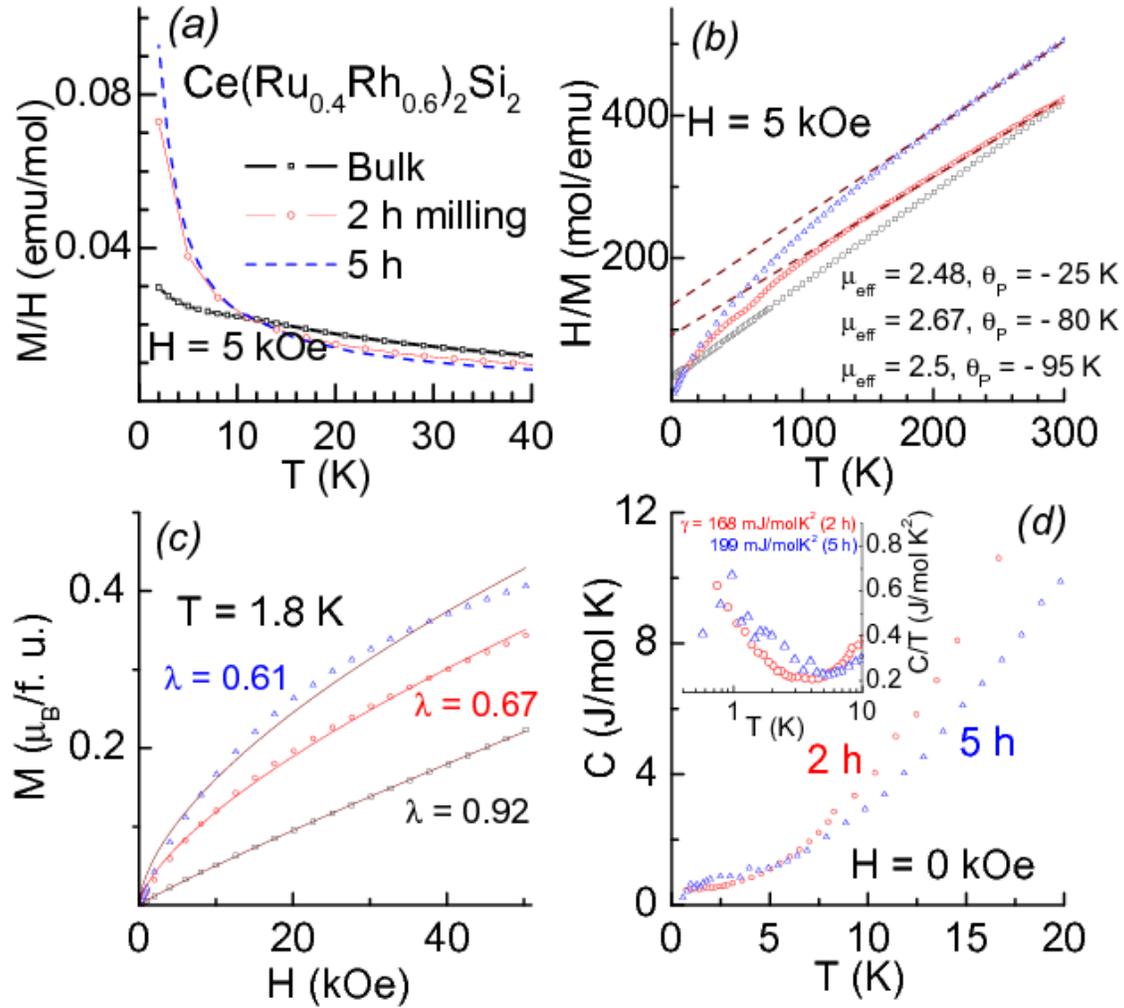

Figure 4:

*(a)* Magnetic susceptibility as a function of temperature, *(b)* inverse susceptibility as a function of temperature, *(c)* isothermal magnetization at 1.8 K and *(d)* heat-capacity as a function of temperature for the bulk and milled specimens of CeRu$_{0.8}$Rh$_{1.2}$Si$_2$. In the inset of *(d)*, heat-capacity divided by temperature is plotted in logarithmic temperature scale for the milled specimens, in addition to showing the linear term obtained from the range 5-10 K. In *(c)*, a fit to H$^\lambda$ is shown by continuous lines. The lines through the data points in *(a)* serve as guides to the eyes and in *(b)* the dashed lines are drawn through high temperature linear region. Magnetic moment (in units of μ$_B$) and paramagnetic Curie temperature are given in *(b)*.

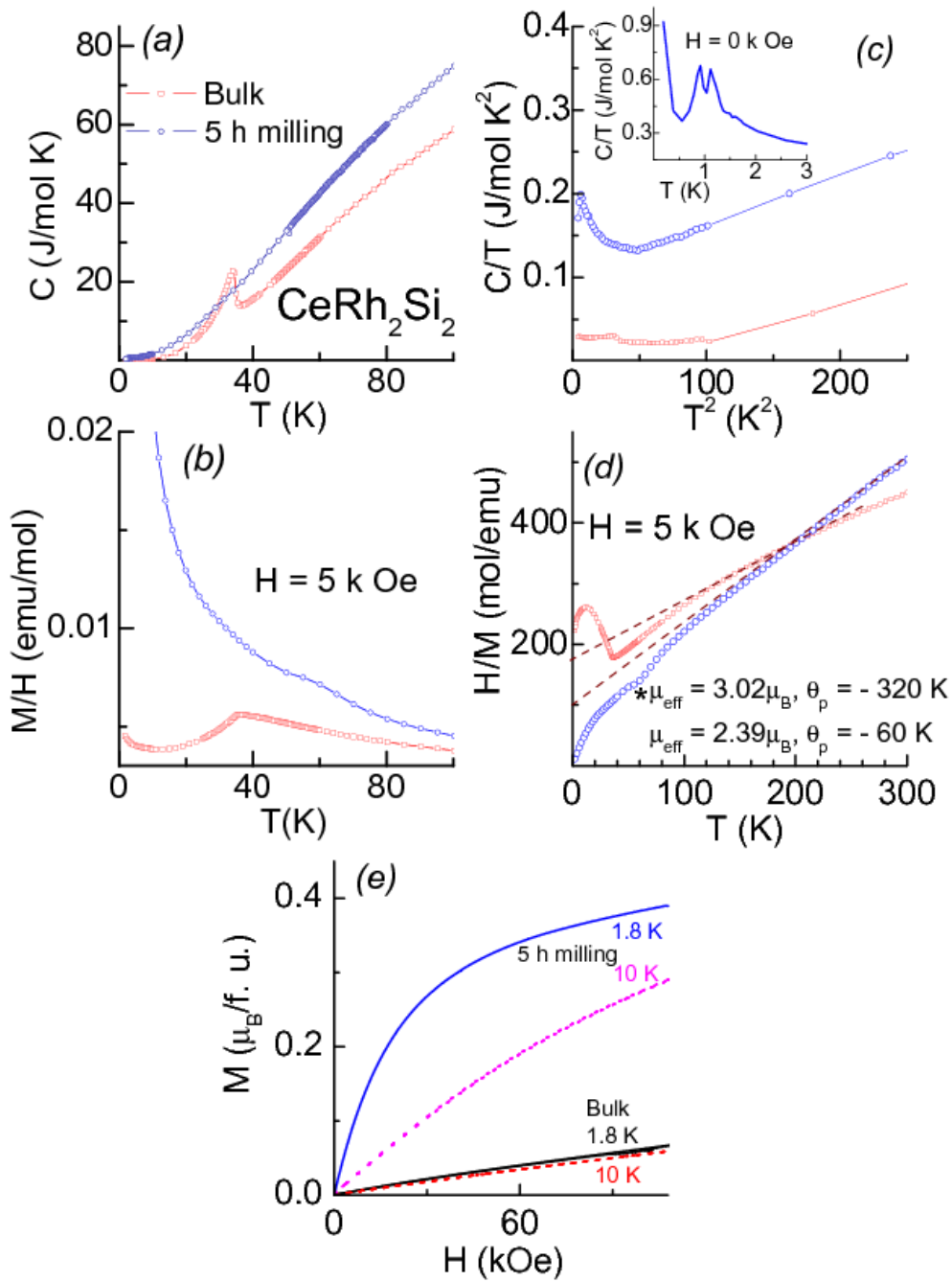

Figure 5:
*(a)* Heat-capacity and *(b)* magnetization divided by magnetic field as a function of temperature for the bulk and milled specimens of $CeRh_2Si_2$. Heat-capacity divided by temperature as a function of square of temperature, inverse susceptibility as a function of temperature, isothermal magnetization as a function of magnetic field are plotted in *(c)*, *(d)*, and *(e)* respectively. In the inset of figure *(c)*, the data below 3 K is plotted to highlight the existence of weak features around 1 K. While the lines through the data points serve as guides to the eyes, the dashed lines in *(d)* are extrapolations of high temperature Curie-Weiss region.